\documentclass[fleqn,usenatbib]{mnras}

\usepackage{newtxtext,newtxmath}

\usepackage[T1]{fontenc}
\usepackage{ae,aecompl}

\usepackage{graphicx}	
\usepackage{amsmath}	
\usepackage{changes}
\usepackage{relsize}

\def\aj{AJ}
\def\apj{ApJ}
\def\apjl{ApJL}
\def\apjs{ApJ Suppl. Ser.}

\def\aap{A\&A}

\def\mnras{MNRAS}

\def\pasj{PASJ}
\def\physrep{Phys. Rep.}

\def\beq#1{\begin{equation}\label{#1}}
\def\eeq{\end{equation}}
\def\beqa#1{\begin{eqnarray}\label{#1}}
\def\eeqa{\end{eqnarray}}

\def\comment#1{\relax}

\title[WR to BH mass connection]{Empirical correlation between masses of black holes and Wolf-Rayet stars derived from their mass distributions in spectroscopic binaries}

\author[K.A. Postnov et al.]{
K.A. Postnov$^{1}$\thanks{E-mail: kpostnov@gmail.com, the corresponding author},
I.A. Shaposhnikov$^{1}$\thanks{E-mail: iv.shaposhnikov@gmail.com}
and
A.M. Cherepashchuk$^{1}$\thanks{E-mail: cherepashchuk@gmail.com}
\\
$^{1}$ Sternberg Astronomical Institute, M.V. Lomonosov Moscow State University, 13, Universitetskij pr., 119234, Moscow, Russia\\
}

\date{Accepted XXX. Received YYY; in original form ZZZ}

\pubyear{2025AB}

\begin{document}
\label{firstpage}
\pagerange{\pageref{firstpage}--\pageref{lastpage}}
\maketitle

\begin{abstract}
We present  {observationally determined} mass distributions of Wolf-Rayet (WR) stars in WR+OB binaries and black holes (BH) in  {spectroscopic} binaries. Both WR and BH mass  {probability} distributions can be well approximated by unbiased log-normal functions.  {Assuming that all WR stars with $M_\mathrm{WR}\gtrsim 6 M_\odot$ after core collapse are progenitors of the BHs,} the similar shape of their mass distributions before and after collapse suggests a power-law relation between them $M_{\mathrm{BH}} \simeq (0.39\pm0.09) {M_{\mathrm{WR}}}^{1.13\pm0.09}$.  {Using the relation between masses of  WR star and its CO-core, we obtain  the empirical relation between the BH mass and CO-core of the collapsing WR star $M_\mathrm{BH}\sim 0.9 M_\mathrm{CO}$, which can be used in population synthesis calculations.}

\end{abstract}




\section{Introduction}

 {Evolutionary scenarios of close binary star (CBS) evolution \citep{1973NPhS..242...71V,1973NInfo..27...70T,2014LRR....17....3P} predict that 
black holes (BHs) in massive binary systems, which are observed as massive or low-mass X-ray binaries (HMXB/LMXB), should have passed through the Wolf-Rayet (WR) stage before the core collapse of the initial primary component}.
Observational data on several dozen WR stars in WR+OB massive binaries (which are thought to be immediate predecessors of HMXBs) and BHs in X-ray binaries enables one to examine their statistical properties, especially their masses. 

BH formation in core collapses remains an open issue, results of numerical simulations are model-dependent \citep{2018ApJ...855L...3O,2025ApJ...987..164B,2025MNRAS.tmp.1413H}.  {While it is theoretically expected that BH formation occurs from the collapses of stellar CO-cores with masses $\gtrsim 6.6 - 8 M_\odot$ \citep{2020A&A...638A..39L,2025MNRAS.tmp.1413H}, the outcome of the core collapse (black hole or neutron star formation) may depend not solely on the CO-core mass \citep{2018ApJ...855L...3O,2025MNRAS.540.3094G}.}
Therefore the WR-BH mass relation derived from empirical data seems to be interesting and potentially can be tested in future model calculations and more detailed observations.
 Assuming that WR stars with masses $\gtrsim 6M_\odot$ and BHs in binaries are generically related, their mass distributions can be used  to search for the empirical transformation law of the WR mass to the BH mass after collapse. This can be done straightforwardly if both distributions are parametrized analytically.

\section{Mass distributions for WR stars and BHs in spectroscopic binaries }

\subsection{Observational data}

Masses of components of binary stars can be most reliably determined by solving spectroscopic orbits with known binary inclination $i$ obtained independently \citep{2022ARep...66S.567C}.

For our analysis, we have used masses of WR stars in WR+OB close binaries in our Galaxy and Magellanic Clouds. The necessary condition to select the binary was the presence of spectral lines from both components (SB2) and the reliable estimate of the binary inclination angle $i$. In addition to these close binaries with WR stars, we have also used the WR star mass estimates from several wide WR+OB binaries obtained from joint solution of the spectroscopic and visual orbits \citep{2024ApJ...977...78R,2024ApJ...977..185H,2021ApJ...908L...3R,2021MNRAS.504.5221T,2007MNRAS.377..415N}. 

\begin{table*}
    \centering
    \begin{tabular}{|c|c|c|c||c|c|c|c|}
        \hline
        Name    & Type           & $M_{\mathrm{WR}},~M_\odot$ & Ref. & Name    & Type           & $M_{\mathrm{WR}},~M_\odot$ & Ref. \\
        \hline
        WR 127	& WN3+O9.5V      & $20.4_{\scriptscriptstyle -2.5}^{\scriptscriptstyle +3.4}$   &  [1,2]    &         WR 35a	& WN6+O8.5V      & $17.7 \pm 1.1$             & [18]            \\
        BAT99 129	& WN3+O5V    & $15.1_{\scriptscriptstyle -0.9}^{\scriptscriptstyle +2.3}$   & [3]       &         WR 153	& WN6+O6I        & $6.41_{\scriptscriptstyle -0.68}^{\scriptscriptstyle +0.91}$    & [19,2]   \\
        WR 151	& WN4+O5V        & $13.9 \pm 0.7$           & [4,5]           &         WR 145	& WN7+O7V(f)     & $17.3_{\scriptscriptstyle -1.1}^{\scriptscriptstyle +1.6}$      & [20]     \\
        BAT99 019	& WN4+O6V        & $22.2 \pm 1.7$             &   [3]     &         Br 22	 & WC4+O5-6V-III  & $11.31_{\scriptscriptstyle -0.23}^{\scriptscriptstyle +1.58}$    & [21]   \\
        SMC AB6	& WN4+O6.5I      & $15.1_{\scriptscriptstyle -1.8}^{\scriptscriptstyle +3.2}$   & [6]       &         WR 9	 & WC4+O7         & $31.7_{\scriptscriptstyle -1.8}^{\scriptscriptstyle +2.5}$     & [21,2]   \\
        SMC AB7 & WN4+O6I(f)     & $24.1 \pm 2.1$             & [7]           &         WR 137	 & WC4+O9e        & $9.49 \pm 3.41$	                   & [22]   \\
        WR 21	& WN5+O4-6       & $16.6_{\scriptscriptstyle -1.4}^{\scriptscriptstyle +2.7}$       & [8,2] &         WR 48	 & WC5+B0III      & $9.05_{\scriptscriptstyle -0.45}^{\scriptscriptstyle +0.91}$     & [23]   \\
        WR 141	& WN5+O5V-III    & $33.9 \pm 3.3$             & [9,2]         &         WR 30	 & WC6+O6-8       & $16.2_{\scriptscriptstyle -0.7}^{\scriptscriptstyle +1.1}$       & [8,2] \\
        WR 62a	& WN5+O5.5-6     & $22.6 \pm 3.6$             & [10]          &         WR 79	 & WC7+O5-8       & $9.8_{\scriptscriptstyle -1.4}^{\scriptscriptstyle +2.3}$        & [23,2] \\
        WR 139	& WN5+O6II-V     & $10.5 \pm 0.5$             & [11,12]       &         WR 140	 & WC7+O5.5fc     & $10.31 \pm 0.45$                     & [24] \\
        WR 138	& WN5+O9V        & $13.9 \pm 1.5$	          & [13]          &         WR 42	 & WC7+O7V        & $13.2_{\scriptscriptstyle -1.4}^{\scriptscriptstyle +2.0}$       & [23,2] \\
        WR 133	& WN5+O9I        & $9.3 \pm 1.6$	          & [14]          &         WR 11	 & WC8+O7.5III-V  & $23.4 \pm 1.4$             & [25]           \\
        WR 68a	& WN6+O5.5-6     & $15.6 \pm 3.5$             & [15]          &         WR 113	 & WC8+O8-9III-V  & $11.7 \pm 0.6$             & [26]           \\
        WR 47	& WN6+O5V        & $15.7 \pm 0.5$             & [8,16]        &         SMC AB 8	 & WO4+O4V        & $13.9_{\scriptscriptstyle -2.7}^{\scriptscriptstyle +4.1}$     & [27] \\
        WR 155	& WN6+O6I        & $10.8 \pm 0.6$             & [4,17]        &         WR 30a	 & WO4+O5((f))    & $9.4_{\scriptscriptstyle -1.4}^{\scriptscriptstyle +3.6}$        & [28]   \\
        \hline
    \end{tabular}
    \caption{Estimates of WR masses $M_{\mathrm{WR}}$. \textbf{References.} [1] \citealt{2024AnA...683L..17S}, [2] \citealt{1996AJ....112.2227L}, [3] \citealt{2019AnA...627A.151S}, [4] \citealt{2023MNRAS.523.1524S}, [5] \citealt{2009PASP..121..708H}, [6] \citealt{2018AnA...616A.103S}, [7] \citealt{2002MNRAS.333..347N}, [8] \citealt{2012MNRAS.424.1601F}, [9] \citealt{2024ARep...68.1145S}, [10] \citealt{2013AnA...552A..22C}, [11] \citealt{2023MNRAS.526.4529S}, [12] \citealt{2011AN....332..616E}, [13] \citealt{2024ApJ...977..185H}, [14] \citealt{2021ApJ...908L...3R}, [15] \citealt{2015AnA...581A..49C}, [16] \citealt{1990ApJ...350..767M}, [17] \citealt{1997AN....318..267D}, [18] \citealt{2014AnA...562A..13G}, [19] \citealt{2002ApJ...577..409D}, [20] \citealt{2009MNRAS.399.1977M}, [21] \citealt{1990ApJ...348..232M}, [22] \citealt{2024ApJ...977...78R}, [23] \citealt{2002MNRAS.335.1069H}, [24] \citealt{2021MNRAS.504.5221T}, [25] \citealt{2007MNRAS.377..415N}, [26] \citealt{2018MNRAS.474.2987H}, [27] \citealt{2016AnA...591A..22S}, [28] \citealt{2001MNRAS.327..435G}. }
    \label{tab:WR}
\end{table*}
\begin{table*}
    \centering
    \begin{tabular}{cccccccc}
        \hline
        Name    & Type           & $M_{\mathrm{BH}},~M_\odot$ & Ref. & Name    & Type           & $M_{\mathrm{BH}},~M_\odot$ & Ref. \\
        \hline
        HD 96670 & HMXB & $6.2 \pm 0.8$ & [1]                       & V821 Ara & LMXB & $5.9 \pm 3.6$ & [26]               \\
        1E 1740.7-2942 & HMXB & $5.1 \pm 0.4$ & [2]                  & V2107 Oph & LMXB & $6.65 \pm 0.25$ & [27]            \\
        SAX J1819.3-2525 & HMXB & $10.2 \pm 0.8$ & [3]               & [KRL2007b] 222 & LMXB & $12.2 \pm 3.5$ & [28]        \\
        SS 433 & HMXB & $9.0_{\scriptscriptstyle \scriptscriptstyle -1.0}^{+3.0}$      &      [4]          & V2293 Oph & LMXB & $6.4 \pm 3.2$ & [29]              \\
        Cyg X-1 & HMXB & $21.2 \pm 2.2$ & [5]                        & XTE J17464-321 & LMXB & $11.2 \pm 1.96$ & [30]       \\
        Cyg X-3 & HMXB & $7.2 \pm 1.0$ & [6]                         & XTE J1752-223 & LMXB & $9.6 \pm 0.9$ & [31]          \\
        M33 X-7 & HMXB & $11.4_{\scriptscriptstyle -1.7}^{\scriptscriptstyle +3.3}$    &  [7]          & 2XMM J180112.4-254436 & LMXB & $6.0 \pm 3.4$ & [32]  \\
        NGC 300 X-1 & HMXB & $17.0 \pm 4.0$ & [8]                    & V4641 Sgr & LMXB & $6.4 \pm 0.6$ & [33]              \\
        IC 10 X-1 & HMXB & $15.0_{\scriptscriptstyle -3.0}^{\scriptscriptstyle +10.0}$         &  [9]   & MAXI J1820+070 & LMXB & $8.48 \pm 0.79$ & [34]       \\
        V518 Per & LMXB & $6.5_{\scriptscriptstyle -2.9}^{\scriptscriptstyle +3.0}$    &      [10]          & MAXI J1836-194 & LMXB & $8.5 \pm 3.5$ & [35]         \\
        1A 0620-00 & LMXB & $5.86 \pm 0.24$ & [11]                    & V406 Vul & LMXB & $7.8 \pm 1.9$ & [36]               \\
        MAXI J0637-430 & LMXB & $5.1 \pm 1.6$ & [12]                  & MAXI J1910-057 & LMXB & $9.98 \pm 3.67$ & [37]       \\
        KV UMa & LMXB & $8.3 \pm 0.28$ & [13]                         & Granat 1915+105 & LMXB & $10.1 \pm 0.6$ & [38]       \\
        GU Mus & LMXB & $11.0 \pm 2.1$ & [14]                         & V1408 Aql & LMXB & $5.0 \pm 1.0$ & [39]              \\
        MAXI J1305-704 & LMXB & $8.9 \pm 1.6$ & [15]                  & QZ Vul & LMXB & $7.2 \pm 1.7$ & [40]                 \\
        MAXI J1348-630 & LMXB & $14.8 \pm 0.9$ & [16]                 & V404 Cyg & LMXB & $10.0 \pm 2.0$ & [41]              \\
        CRTS J135716.8-093238 & LMXB & $12.4 \pm 3.6$ & [17]          & MM Vel & LMXB & $4.4 \pm 0.5$ & [42]                 \\
        KY TrA & LMXB & $5.8 \pm 3.0$ & [18]                          & MAXI J1728-360 & LMXB & $4.6 \pm 0.5$ & [43]         \\
        MAXI J1535-571 & LMXB & $10.39 \pm 0.62$ & [19]               & XTE J1818-245 & LMXB & $4.0 \pm 0.5$ & [44]          \\
        MAXI J1543-564 & LMXB & $13.0 \pm 1.0$ & [20]                 & MAXI J1828-249 & LMXB & $4.0 \pm 0.5$ & [45]         \\
        IL Lup & LMXB & $5.0 \pm 2.5$ & [21]                          & EXO 1846-031 & LMXB & $3.24 \pm 0.2$ & [46]          \\
        V381 Nor & LMXB & $9.4 \pm 1.4$ & [22]                        & \textit{Gaia} BH1 & Wide bin. & $9.62 \pm 0.18$ & [47]       \\
        X Nor X-1 & LMXB & $10.0 \pm 0.1$ & [23]                      & \textit{Gaia} BH2 & Wide bin. & $8.94 \pm 0.34$ & [48]        \\
        V1033 Sco & LMXB & $5.31 \pm 0.07$ & [24]                    & \textit{Gaia} BH3 & Wide bin. & $32.7 \pm 0.82$ & [49]        \\
        MAXI J1659-152 & LMXB & $5.4 \pm 2.1$ & [25]                  & OGLE-2011-BLG-0462 & Single & $7.15 \pm 0.83$ & [50] \\
        \hline
    \end{tabular}
    \caption{Estimates of BH masses $M_{\mathrm{BH}}$. \textbf{References.} 
    [1] \citealt{2021ApJ...913...48G}, [2] \citealt{2020MNRAS.493.2694S}, 
    [3] \citealt{2001ApJ...555..489O}, [4] \citealt{2023NewA..10302060C}, 
    [5] \citealt{2022ApJ...934....4K}, [6] \citealt{2022ApJ...926..123A}, 
    [7] \citealt{2022AnA...667A..77R}, [8] \citealt{2021ApJ...910...74B}, 
    [9] \citealt{2024ApJ...974..184W}, [10] \citealt{2024MNRAS.531.4917C}, 
    [11] \citealt{2017MNRAS.472.1907V}, [12] \citealt{2022MNRAS.515.3105S}, 
    [13] \citealt{2012ApJ...744L..25G}, [14] \citealt{2016ApJ...825...46W}, 
    [15] \citealt{2021MNRAS.506..581M}, [16] \citealt{2023AnA...669A..57T}, 
    [17] \citealt{2016ApJ...822...99C}, [18] \citealt{2024MNRAS.527.5949Y}, 
    [19] \citealt{2019MNRAS.487.4221S}, [20] \citealt{2016ApJ...827...88C}, 
    [21] \citealt{1998ApJ...499..375O}, [22] \citealt{2002ApJ...568..845O}, 
    [23] \citealt{2014ApJ...789...57S}, [24] \citealt{2014MNRAS.437.2554M}, 
    [25] \citealt{2021MNRAS.501.2174T}, [26] \citealt{2017ApJ...846..132H}, 
    [27] \citealt{1997PASP..109..461F}, [28] \citealt{2015ApJ...807..108I}, 
    [29] \citealt{2023MNRAS.526.5209C}, [30] \citealt{2017ApJ...834...88M}, 
    [31] \citealt{2010ApJ...723.1817S}, [32] \citealt{2006PhDT........26B}, 
    [33] \citealt{2014ApJ...784....2M}, [34] \citealt{2020ApJ...893L..37T}, 
    [35] \citealt{2014MNRAS.439.1381R}, [36] \citealt{2022MNRAS.517.1476Y}, 
    [37] \citealt{2023AdSpR..71.1045N}, [38] \citealt{2013ApJ...768..185S}, 
    [39] \citealt{2021RAA....21..214S}, [40] \citealt{2004AJ....127..481I}, 
    [41] \citealt{1992ApJ...401L..97W}, [42] \citealt{2003AnA...404..301R}, 
    [43] \citealt{2023MNRAS.519..519S}, [44] \citealt{2021ApJ...912..110B}, 
    [45] \citealt{2019PASJ...71..108O}, [46] \citealt{2020AAS...23515902S}, 
    [47] \citealt{2023MNRAS.518.1057E}, [48] \citealt{2023MNRAS.521.4323E}, 
    [49] \citealt{2024AnA...686L...2G}, [50] \citealt{2025ApJ...983..104S}.
    }
    \label{tab:BH}
\end{table*}
\begin{figure*}
    \centering \includegraphics[width=0.95\linewidth]{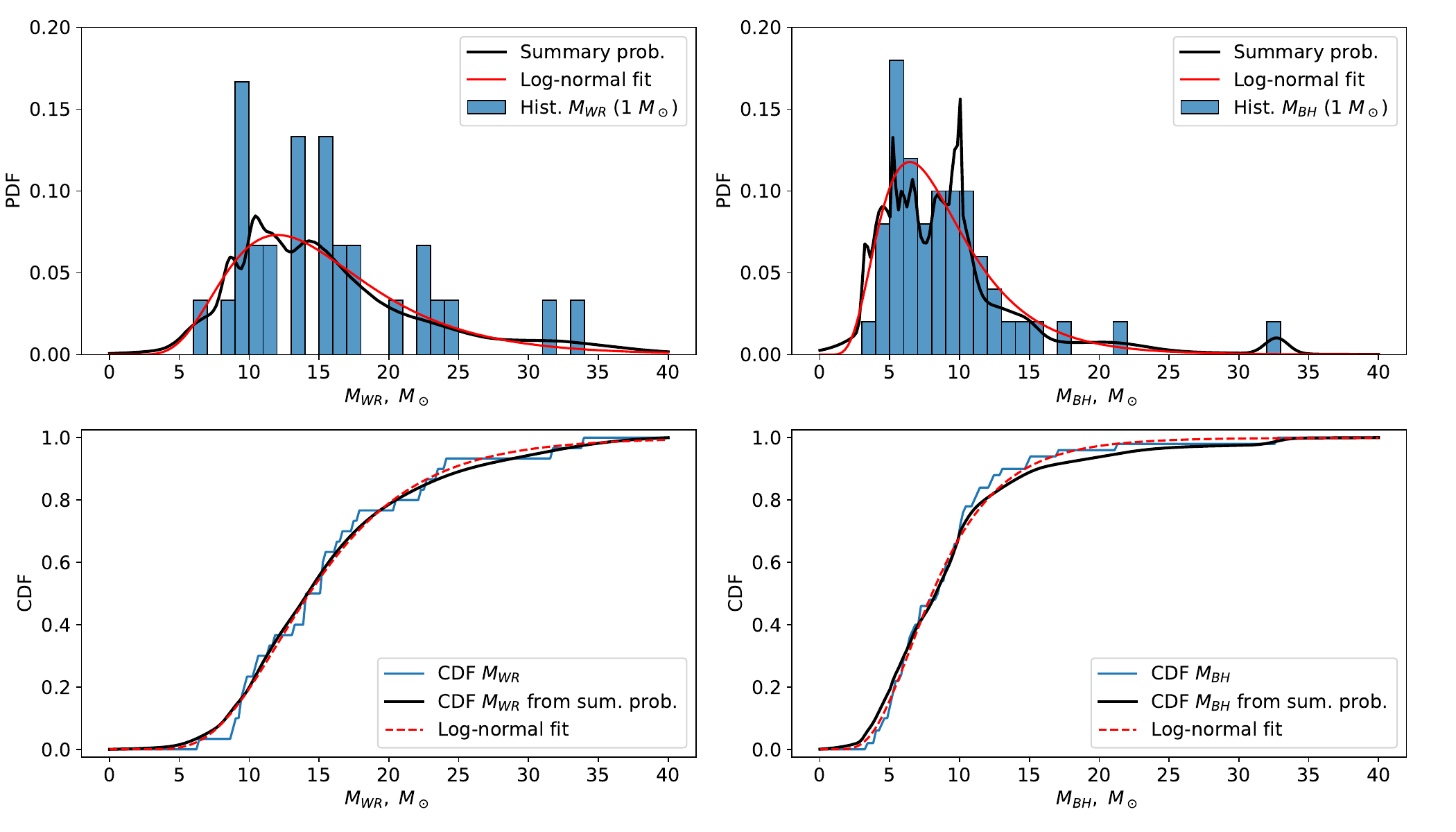}
    \caption{Differential (upper panels) and cumulative (bottom panels) WR  {(to the left)} and BH  {(to the right)} mass distributions based on data from Tables \ref{tab:WR} and \ref{tab:BH}. In red shown in the best log-normal fit (\textit{mean} in Table 3).}
    \label{fig:WR-BH}
\end{figure*}

Masses of compact objects in X-ray binaries were taken from recent HMXB and LMXB catalogs
\citep{2023AnA...671A.149F,2024AnA...684A.124F,2023AnA...677A.134N,2023AnA...675A.199A} and some later papers \citep{2023NewA..10302060C,2024MNRAS.531.4917C}. We have also included three BH candidates from \textit{Gaia} wide binaries \citep{2023MNRAS.518.1057E,2023MNRAS.521.4323E,2024AnA...686L...2G} and the mass of single BH candidate OGLE-2011-BLG-0462 obtained from gravitational microlensing \citep{2025ApJ...983..104S}. Note that we have not included BH mass estimates in coalescing binaries found from gravitational wave measurements \citep{2025arXiv250818082T} because these BHs may result from different  formation channels, such as dynamical captures in dense stellar clusters \citep{PhysRevD.100.043027}, collapses of Population III stars \citep{2016MNRAS.456.1093K}, primordial BHs \citep{2016JCAP...11..036B}, etc. (see, e.g., \citealt{2022PhR...955....1M} for more detail and references).

\subsection{WR and BH mass estimates}

We took WR mass estimates $M_{\mathrm{WR}}$ in SB2 spectroscopic binaries as derived from the observed values $M_\mathrm{obs} \equiv M_{\mathrm{WR}} \sin^3 i$, the binary inclination $i$ and their determination errors reported in the literature. The individual WR mass probability distribution was calculated by bootstrapping of $M_\mathrm{obs}$, and $i$ within their errors (Gaussian or uniform, see \citealt{2014ARep...58..113P} for more detail on the method of the individual mass probability distribution determination).  

To estimate $M_{\mathrm{BH}}$ in  X-ray binaries which are SB1 spectral binary systems with one unseen companion, three observables are required: the optical star mass function $f_v \equiv f(M_v)$, the mass ratio of the components $q \equiv M_v/M_{\mathrm{BH}}$, and the binary inclination angle $i$. Then the BH mass is estimated as $M_{\mathrm{BH}} = f_v (1+q)^2 (\sin{i})^{-3}$ and its probability distribution is calculated by bootstrapping of the parameters within their observational errors.

By summing up individual probability distributions for WR-star masses  ($j=[1,...,N_{\mathrm{WR}}]$, $N_{\mathrm{WR}} = 30$ is the total number of WR+OB in our sample) and individual distributions of BH masses $j$ ($j=[1,...,N_{\mathrm{BH}}]$, $N_{\mathrm{BH}} = 50$ is the total number of BH CBS in our sample)  we construct a continuous full mass probability distribution \citep{2010ApJ...725.1918O,2014ARep...58..113P}: 
\begin{equation}
    f_\mathrm{type}(M) = \frac{1}{N_\mathrm{type}}\sum_{j=1}^{N_\mathrm{type}} f^{(j)}_\mathrm{type}(M),\quad \mathrm{type}=[\mathrm{WR},\mathrm{BH}]  \,. 
\end{equation}

The WR and BH mass estimates $M_{\mathrm{BH}}$ and references are listed in Tables \ref{tab:WR}-\ref{tab:BH}. Fig. \ref{fig:WR-BH} shows the WR (left panels) and BH (right panels) mass distributions in the form of histograms with 1 $M_\odot$ bins (upper panels) and cumulative distribution functions (bottom panels). The most probable mass values are in blue, the continuous estimates of distributions are shown by black curves, log-normal fits (see below) are shown by red curves. 

\subsection{Log-normal approximation}

It is possible to describe the obtained WR and BH mass distributions analytically. The best-fit is obtained by a log-normal function
\begin{equation}
\label{e:lognorm}
    f(M)dM=\frac{1}{M\sqrt{2\pi}s}\exp\left(-\frac{\ln^2(M/M_0)}{2s^2}\right)dM\,.
\end{equation}
The parameters $M_0,s$ were found from fitting the most probable WR and BH masses (in blue in Fig. \ref{fig:WR-BH}) and from the continuous sum of individual probability distributions (black curves in Fig. \ref{fig:WR-BH}). The optimal parameters where chosen as the mean values from the two fits. 
The obtained parameters of the log-normal fits are listed in Table \ref{tab:lognorm}. Log-normal WR and BH mass distributions with optimal parameters $M_0,s$ are shown by red curves in Fig. \ref{fig:WR-BH}.

\begin{table}
    \centering
    \begin{tabular}{ccc}
        \hline
        Dist. to fit & $s$ & $M_{0},~M_\odot$ \\
        \hline
        WR \textit{hist.} & 0.39 & 14.40 \\
        WR \textit{curve} & 0.44 & 14.18 \\
        WR \textit{mean}  & $0.42\pm0.02$ & $14.29\pm0.11$\\
        \hline
        BH \textit{hist}. & 0.45 & 8.07 \\
        BH \textit{curve} & 0.50 & 8.00 \\
        BH \textit{mean}  & $0.47\pm0.03$ & $8.03\pm0.03$ \\
        \hline
    \end{tabular}
    \caption{Parameters of the log-normal distribution (2) obtained from fitting the histogram (\emph{hist.}) and the sum of individual probabilities (\emph{curve}), and their mean value (\emph{mean}, log-normal fit in red in Fig. \ref{fig:WR-BH}).}
    \label{tab:lognorm}
\end{table}

\section{Discussion}

That both mass distributions of WR-stars (assumed progenitors of compact stars) and BHs in CBS have similar shape enables us to make conclusion on the WR mass change during the compact object formation. It is straightforward to show that if values $x$ and $y$ are distributed log-normally (\ref{e:lognorm}) with parameters $M_0 = M_1, s = s_1$ and  $M_0 = M_2, s = s_2$, respectively, the transformation law $x\to y$ reads
\begin{equation}
    y = M_2 \left( \frac{x}{M_1} \right)^{\frac{s_2}{s_1}} \equiv C x^p \quad (p = s_2/s_1,~C = M_2 M_1^{-p})\,.
\end{equation}
In our case of $M_{\mathrm{WR}}$ and $M_{\mathrm{BH}}$ distributions with parameters from Table \ref{tab:lognorm} we get 
\begin{equation}
\label{e:mbhmwr}
    M_{\mathrm{BH}} \simeq (0.39\pm0.09) {M_{\mathrm{WR}}}^{1.13\pm0.09} 
\end{equation}

It is interesting to compare the obtained power-law WR-BH mass relation (\ref{e:mbhmwr}) with known relation between masses of He stars with their CO-cores \citep{1971AcA....21....1P}: $M_\mathrm{CO}\approx 0.45 {M_\mathrm{WR}}^{1.2}$. 
Note that the power law index in Eq. (\ref{e:mbhmwr}) derived from empirical distributions is close to that in ${M_\mathrm{CO}}-M_\mathrm{WR}$ relation found from calculations of evolution of massive stars\footnote{The results of recent calculations \cite{2023ApJ...945...19T} give a similar dependence $M_\mathrm{CO} \approx  0.38 {M_\mathrm{He}}^{1.25}$ for $M_\mathrm{He}>5 M_\odot$ (see their Fig. 1).}. 
Using Eq. (\ref{e:mbhmwr}) we obtain the relation between the masses of CO-cores and BHs:
\begin{equation}
\label{e:mbhmco_1}
    M_{\mathrm{BH}}\simeq (0.82\pm0.19) {M_{\mathrm{CO}}}^{0.94\pm0.08}\,.
\end{equation}
This relation, however, ignores the stellar wind mass loss of WR stars, which can be significant. Let us obtain relation (\ref{e:mbhmco_1}) with taking into account stellar wind mass loss in the limiting case where the observed WR mass distribution is assumed to correspond to WR-stars at the beginning of the WR stage.

In the power-law approximation of the stellar-wind mass loss \citep{1989AnA...220..135L,2000AnA...360..227N,2017AnA...607L...8V} $\dot M_{\mathrm{WR}} = k M_{\mathrm{WR}}^\alpha$ ($k<0$), at the end of WR stage the final WR mass $M_{\mathrm{WR},f}$ is related to the initial one $M_{\mathrm{WR},i}$ as
\begin{equation}
    {M_{\mathrm{WR},f}}^{1-\alpha} = {M_{\mathrm{WR},i}}^{1-\alpha} + k(1-\alpha)T\,.
\end{equation}
Assume that the WR lifetime is $T = T(M_{\mathrm{WR},f}) = C_T / \sqrt{M_{\mathrm{WR},f}}$, $C_T \approx 1.74\cdot10^6~[\mathrm{yr}]\cdot M_\odot^{1/2}$ \citep{2022ARep...66S.567C}, and $\alpha = 1.5$, $k =-2\cdot10^{-7}~[\mathrm{yr}]^{-1}\cdot M_\odot^{-1/2}$ (cf. with the empirical estimate $\dot M_{WR} \simeq -(1.6\pm0.7)\cdot10^{-7}{M_\mathrm{WR}}^{1.61\pm0.29}$, as inferred from observations of secular orbital period increase in WR+OB stars in papers  \citealt{2024maeu.conf..550S,2024ARep...68.1145S}). It is easy to see that in the case $\alpha=1.5$ the final WR mass is a fraction of the initial one:
\begin{equation}
    M_{\mathrm{WR},f} = (1-0.5|k|C_T)^2\cdot M_{\mathrm{WR},i} \approx 0.7 M_{\mathrm{WR},i}.
\end{equation}
Then in relation (\ref{e:mbhmco_1}) the power-law index remains the same and only the proportionality coefficient is slightly modified:
\begin{equation}
\label{e:mbhmco_2}
    M_{\mathrm{BH}}\simeq (1.19\pm0.28){M_{\mathrm{CO}}}^{0.94\pm0.08}\,.
\end{equation}
Thus,  {in both cases -- without mass-loss and with taking into account the WR mass decrease due to stellar-wind mass loss over the entire life of WR star --} the BH mass can be close to the CO-core mass of the collapsing star, $M_\mathrm{BH}\sim  0.9 M_\mathrm{CO}$ -- the BH formation prescription  sometimes used in population synthesis calculations (e.g. \citealt{2019MNRAS.483.3288P}).

We emphasize that our main relation (4) between the WR and BH mass in spectroscopic binaries was derived under one assumption that the WR-star stage precedes the formation of BH in core collapse. We took into account only massive WR-stars $\gtrsim 6 M_\odot$, although close to this lower mass limit the collapse of CO-core of WR-star can lead, according to some model calculations \citep{2018ApJ...855L...3O,2025MNRAS.540.3094G}, to neutron star formation, which we do not consider in the present study. We have also ignored possible change in BH mass due to mass transfer in binaries due to the short life-time of the secondary star in massive binaries or low accretion rates in low-mass binaries. 

\section{Conclusion}

We have constructed empirical mass distributions of WR-stars in  {observed} WR+OB binaries in which the WR mass determination is most reliable (Table 1). We have also constructed the mass distribution of BHs in HMXB and LMXB, added with three BHs in \textit{Gaia} wide binaries and one single BH candidate from microlensing observations (Table 2). Both mass distributions were successfully fitted by log-normal functions. Assuming that WR-star stage precedes the BH formation in binaries, the transformation of WR to BH mass distribution enabled us to found the connection between the WR and BH mass after the core collapse (Eq. \ref{e:mbhmwr}). This relation derived from empirical mass distribution is similar to that of WR mass to its CO-core derived from massive star evolutionary calculations \citep{1971AcA....21....1P,2023ApJ...945...19T}. This result may justify the use of simple relation between the mass of BH and CO-core of collapsing star, $M_\mathrm{BH}\approx 0.9 M_\mathrm{CO}$. It would be important to test our findings on larger statistics of WR and BH masses in binaries. 

\section*{Acknowledgements}

We thank the anonymous referee for useful notes that helped us to improve the presentation of our results.
The work of IASh and AMCh is supported by the Russian Science Foundation through grant 23-12-00092. 

\section*{Data availability}
Data are available from the authors by request.





\bsp	
\label{lastpage}
\end{document}